\documentclass[showpacs,aps,prb,twocolumn,floatfix]{revtex4}
\usepackage{graphicx}

\begin{document}

\title{Transport through an interacting system connected to leads}

\author{G. Chiappe}
\email{gchiappe@dfuba.df.uba.ar}

\affiliation{Departamento de F\'{\i}sica J. J. Giambiagi,
Facultad de Ciencias Exactas,
Universidad de Buenos Aires, Ciudad Universitaria,
1428 Buenos Aires, Argentina.}

\author{J. A. Verg\'es}
\email{jav@icmm.csic.es}

\affiliation{Instituto de Ciencia de Materiales de Madrid,
Consejo Superior de Investigaciones Cient\'{\i}ficas,
Cantoblanco, E-28049 Madrid, Spain.}

\date{Received on 29 May 2003}

\begin{abstract}
Keldysh formalism is used to get the current-voltage characteristic of
a small system of interacting electrons described by a Hubbard model
coupled to metallic wires. The numerical procedure is checked recovering
well--known results for an Anderson impurity. When larger interacting regions
are considered quite different results are obtained depending on whether
the Hubbard part is half--filled or not.
At half--filling the existence of an energy gap for charge excitations
manifests itself making
current exponentially small as a function both of the number of interacting
sites and the value of $U$. The behavior changes at large voltages
above the gap energy when activated charge transport takes place.
On the contrary, for filling factors other than half,
current goes through the interacting system suffering just a small amount
of scattering at both connections. Conductance depends slightly
on $U$ and much more on the filling factor but not on the length of
the interacting region.
\end{abstract}

\pacs{73.23.-b, 73.63.Nm, 72.10.Bg, 71.10.Fd}

\maketitle

\section{Introduction}

There are several reasons for studying the precise role played by
electron--electron interactions in the transport of electric current
through small systems connected to metallic electrodes. 
Among others we can mention quantum dot measurements in the Kondo
regime, deviations of the exact conductance quantization in low--disorder
quantum wires, possibility of many--body effects in the transport through
molecules, or even, quite fundamental discussions about the possibility
of quantum dephasing at $T=0$.
Nevertheless, in this contribution we are not particularly concerned
with the explanation of any detailed experimental result but with
the methodological improvement of the numerical tools that can be used
when studying many--body effects on the electric transport in small
(either mesoscopic or nanoscopic) devices.
For this purpose, we take the simplest interacting model
(the one--dimensional Hubbard model)
connected to metallic leads and develop a precise
numerical recipe to obtain results that make physical sense.
Let us refer as direct precursor of our work the contributions by
Maslov and Stone\cite{maslov}, Ponomarenko\cite{pono} and
Schulz and Safi\cite{schulz} showing that electrons can traverse a
Luttinger liquid without suffering scattering.
Of particular importance are the works by
Oguri\cite{oguri,oguri2} which has studied the same model using exact
second--order perturbation theory in the interaction
concluding that not only transport through a single interacting site
shows perfect transmission at half--filling (symmetric Anderson impurity)
but transport through {\it any} odd number of interacting sites.
We will show that our results point to a regular odd/even behavior
leading to isolating transport characteristics well described by the
existence of a finite energy gap for charge excitations. We will add
physical arguments to our numerical results (see subsection IIIB).
Additional theoretical contributions about transport through 1D Hubbard model
are collected in Ref.(\onlinecite{ymas}).
The theoretical method is presented in the next Section while results
are given in Section III. 
After checking the numerical method for only one interacting site
(Anderson impurity), the conductivity of extended
interacting systems has been studied both at and below half--filling.
The last Section summarizes our findings.

\section{Theoretical Method}

The standard setup of transport measurements in mesoscopic systems is
considered. Ideal metallic leads are connected to a central part containing
the relevant part of the system: interacting electrons in our case.
The evaluation of the current--voltage (IV) characteristic of the extended
system proceeds in two steps. First, the central interacting part is
exactly solved for a given number of electrons and well--defined
electrochemical potential profile by means
of Lanczos type techniques. Second, Keldysh formalism is used to
attach {\em in an approximate but physically appealing way}
the leads to the central region.
Caroli {\em et al.}\cite{caroli} were probably among the first
authors using this kind of direct
approach for the calculation of the tunneling current through a
non--interacting system. The extension of the formalism
for the current through an interacting electron region was given
by Meir and Wingreen\cite{meir}. The formalism depends crucially on
the knowledge of the many--body selfenergy describing interaction effects.
Our numerical procedure gives a precise recipe for an approximate
evaluation of this selfenergy.

\subsection{Finite system}

The central region is described by a nearest--neighbors tight--binding
Hamiltonian with on--site interaction terms among electrons of different
spin. This is the well--known Hubbard model:

\begin{equation}
{\widehat {\cal H}} = t \sum_{<i,j>,\sigma} 
                     {\hat c}^{\dag}_{i,\sigma} {\hat c}_{j,\sigma}
+ U \sum_{i} {\hat n}_{i,-\sigma} {\hat n}_{i,\sigma} \qquad ,
\label{hubbard}
\end{equation}

\noindent
where the label $i$ denotes a generic site in the finite cluster,
$<i,j>$ represents nearest--neighbors pairs, and
${\hat n}_{i,\sigma} = {\hat c}^{\dag}_{i,\sigma} {\hat c}_{i,\sigma}$
gives the number of electrons of spin $\sigma$ at site $i$.

The presence of an external electrostatic potential is described
by an additional one--electron term:

\begin{equation}
{\widehat {\cal H}_{ext}} = -e
\sum_{i,\sigma} \phi_i {\hat n}_{i,\sigma} \qquad ,
\label{perfil}
\end{equation}

\noindent
being $\phi_i$ the value of electrostatic potential at site $i$.
Lanczos method allows the calculation of the groundstate of the finite
interacting system for a number of sites equal or smaller than fourteen
at half--filling
(computation is sometimes feasible for larger number of sites when
a smaller number of electrons is considered or
system symmetries are explicitly used).
Once the groundstate is known, a continued fraction expansion
of the Green functions in a Lanczos type basis allows their evaluation
as a function of energy
(see the review by Dagotto\cite{dagotto} for a detailed description).
In fact, the four Green
functions entering into Keldysh formalism are explicitly calculated in this
way\cite{keldysh,landau,mahan}.
A matrix form with site and Keldysh indices will be used for the whole
set of Green functions of the finite cluster\cite{aclaracion}:

\begin{equation}
\mathbf{g}= \left( \begin{array}{cc}
\mathrm{g^{--}} & -\mathrm{g^{-+}} \\
\mathrm{g^{+-}} & -\mathrm{g^{++}}
\end{array} \right)  \qquad .
\label{greenmin}
\end{equation}

\noindent
Notice that the $\omega$--dependence of Green functions is not shown
in order to keep notation clearer.

\subsection{Coupled system}

Green functions of the extended system (interacting central part plus
electrodes) are obtained by means of a Dyson equation. Leads are described
by a constant density of states equal to the one at the bandcenter
of a non--interacting chain with a hopping equal to $t$.
For example, the Green function at the right
end (site $l$) of the left electrode is:

\begin{equation}
t \; {\mathbf g}{_{l,l}} = \left( \begin{array}{cc}
i & -2i \\
0 & -i
\end{array} \right) \qquad ,
\label{leadocc}
\end{equation}

\noindent
for energies below the chemical potential $\mu_l$ of this electrode and:

\begin{equation}
t \; {\mathbf g}{_{l,l}} = \left( \begin{array}{cc}
-i  & 0 \\
-2i & i
\end{array} \right) \qquad ,
\label{leadvac}
\end{equation}

\noindent
for energies above $\mu_l$. Dyson equation is used to couple
the central part to the leads {\it when} $U=0$:

\begin{equation}
\mathbf{G_0}= \mathbf{g} (0) + \mathbf{g} (0) \mathbf{T} \mathbf{G_0} \qquad ,
\label{keldysh}
\end{equation}

\noindent
being the only non--zero elements of the hopping term $\mathbf{T}$ the
following:

\begin{equation}
{\mathrm{T}}^{--}_{1,l} =
{\mathrm{T}}^{--}_{l,1} =
{\mathrm{T}}^{++}_{1,l} =
{\mathrm{T}}^{++}_{l,1} = t_l \qquad ,
\label{hopleft}
\end{equation}

\noindent
which couple the first site $1$ of the finite system to the
left electrode and

\begin{equation}
{\mathrm{T}}^{--}_{L,r} =
{\mathrm{T}}^{--}_{r,L} =
{\mathrm{T}}^{++}_{L,r} =
{\mathrm{T}}^{++}_{r,L} = t_r  \qquad ,
\label{hopright}
\end{equation}

\noindent
coupling the last site $L$ of the central part to the right electrode.
In this way, the non--interacting
non--equilibrium Keldysh Green functions of the extended system
$\mathbf{G_0}$ can be obtained from the $U=0$ solution of the central
part $\mathbf{g} (0)$. Many--body interactions are added to $\mathbf{G_0}$
by means of a second Dyson equation:

\begin{equation}
\mathbf{G}= \mathbf{G_0} + \mathbf{G_0} \mathbf{\Sigma} \mathbf{G}  \qquad ,
\label{dyson}
\end{equation}

\noindent
where $\mathbf{\Sigma}$ contains all contributions coming from the
Hubbard interaction in (\ref{hubbard}). It is numerically obtained
as the difference between the non--interacting Green function inverse
and the one including $U$ effects:

\begin{equation}
\mathbf{\Sigma}= [\mathbf{g} (0)]^{-1} - [\mathbf{g} (U)]^{-1}  \qquad .
\label{dyson2}
\end{equation}

\noindent
This equation contains all the
approximations needed to get the transport properties of
the extended many--body system. They are:
\newline
i) although $\mathbf{\Sigma}$ is spatially localized (remember that
interaction is restricted to the central part), correlation among
electrons visiting distant parts of electrodes is not taken into account.
\newline
ii) the detailed form of the electrostatic potential inside the sample
is not known.
\newline
iii) averages within the central part are evaluated using the equilibrium
groundstate of the cluster. They may be changed by the presence of
an important potential bias inducing a large current through the sample.

\noindent
Notice the quite different character of the three approximations
we are using to get well--defined
transport properties through the interacting system.
The first one is just a computational
limitation that can be eventually overcome with larger mainframes
allowing the inclusion of part of the electrodes until convergence
of the selfenergy could been reached.
The second one is physically more relevant and has to do with Hartree
consistency along the system. Although a charge--dependent term could
be added to Eq. (\ref{hubbard}) and iteratively solved, we think that given
the unrealistic one--dimensional nature
of the model this could lead to numerical
difficulties not directly related to physical properties. Therefore, we
have preferred the inclusion of an arbitrary fixed form for the
electrochemical potential inside the sample and just checked that
our main results do not depend qualitatively on the selected profile.
The third one is actually {\it the} approximation of this method.
Naively, one can think that the many--body state of the interacting
part when a current is flowing is not related to any state of the
isolated system nor even to an easy combination of the groundstate
with some excited states. Since the selfenergy defined by Eq. (\ref{dyson2})
directly depends on the state used to calculate Green functions,
one should conclude that this is just an uncontrolled approximation that
has to be validated {\it a posteriori} once results make sense.

Alternatively, many--body contributions to the selfenergy can be
calculated by perturbation theory. Typically, just the second order
diagram is considered\cite{oguri,c5}.

Simultaneous use of equations (\ref{keldysh}), (\ref{dyson}) and
(\ref{dyson2}) gives the main working equation of this paper:

\begin{equation}
\mathbf{G}= \mathbf{g} (U) + \mathbf{g} (U) \mathbf{T} \mathbf{G}  \qquad ,
\label{keldysh2}
\end{equation}

\noindent
where $\mathbf{G}$ is an approximate Keldysh Green function of the
non--equilibrium many--body extended system.
Eq. (\ref{keldysh2}) has been recently used for the study of
Bohm-Aharonov and persistent
currents through quantum dots in the Kondo regime and the
transport in coupled quantum dots\cite{brasil}.
Seminal works using similar ideas to gather exact solutions of small
many--body systems into a lattice are much older\cite{seminal}.

The explicit form of the global Keldysh Green function is:

\begin{equation}
\mathbf{G}= \left( \begin{array}{cc}
\mathrm{G^{--}} & -\mathrm{G^{-+}} \\
\mathrm{G^{+-}} & -\mathrm{G^{++}}
\end{array} \right) \qquad .
\label{green}
\end{equation}

\noindent
Any interesting physical magnitudes can be obtained from it.
For example, the density of states at site $i$ is:

\begin{equation}
N_i(\omega)=-{1 \over  \pi} {Im(
            {\mathrm G}^{--}_{i,i}-{\mathrm G}^{-+}_{i,i}   )}  \qquad ,
\label{dos}
\end{equation}

\noindent
assuming spin degeneracy. If not the case, the spin label should be
explicitly considered doubling the size of the Green function matrices.
Apart from this technical complication, the spin--dependent
formalism remains exactly the same.

The occupied part is directly taken into account by the non--diagonal
part of the Keldysh Green function formalism. Therefore, the charge on
site $i$ is given by:

\begin{equation}
q_i= {e \over {2 \pi}} \int_{-\infty}^{+\infty} d\omega \;
               {Im(  {\mathrm G}^{-+}_{i,i}  )}  \qquad .
\label{den}
\end{equation}

In the same way, the current flowing from left electrode to the first
site in the central cluster labeled by $1$ is:

\begin{equation}
I_{l,1}={{e \, t_l} \over h} \; \int_{-\infty}^{+\infty} d\omega \;
        ({\mathrm G}^{-+}_{1,l} - {\mathrm G}^{-+}_{l,1})  \qquad .
\label{cur}
\end{equation}

\noindent
Differential conductance is obtained as the derivative of the current
versus voltage curve:

\begin{equation}
G= {{d I} \over {d V}} \qquad ,
\label{conductance}
\end{equation}

\noindent
and will be plotted as a function of the energy window $e V$ opened 
by the source--drain voltage $V$. In the particular case in which voltage
does not change the electrostatic potential inside the interacting region,
$G$ will coincide with the function inside integral (\ref{cur}). This
magnitude gives (apart from a constant) the transmission as a function of
band energy for a one--electron
problem but differs from it in a general case due to the inclusion of
inelastic effects. At zero bias, differential conductance and conductance
are equal and can be straightforwardly compared with conductance values
obtained within linear response theory:

\begin{equation}
G_{dc}=t_l \; 
       ({\mathrm G}^{-+}_{1,l} (\mu) - {\mathrm G}^{-+}_{l,1} (\mu))  \qquad ,
\label{dc}
\end{equation}

where Green functions are evaluated at the equilibrium
chemical potential energy $\mu$.

\section{Results}

The numerical method described in the previous section is still not fully
determined. Before giving the results that we believe better describe
the physical problem, we have to give the details that allow the
calculation of a unique current--voltage (IV) characteristic for the problem.

\begin{figure}
\includegraphics[width=0.45\textwidth]{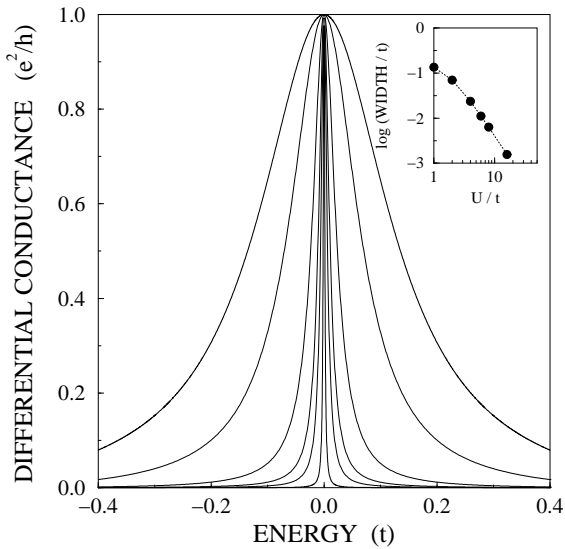}
\caption{Differential conductance per spin channel
through a single Anderson impurity in a half--filled situation
plotted as a function of energy window $e V$ for
several $U$ values ($U/t=1,2,4,6,8,16$). The inset shows the resonance width
as a function of $U/t$ in a log--log representation.}
\label{kondo2.ps}
\end{figure}

The first issue is the value of the electrostatic potential entering
into the central part\cite{c5bis}.
Ideally, one could use Poisson equation to get
the potential from a knowledge of the charge distribution. This would
imply a self--consistent calculation involving the whole system formed
by electrodes and interacting part.
As said before, the one--dimensional character
of the problem adds both technical and physical complications to the
problem. As a consequence, we have preferred the use of a model potential
profile that is numerically convenient and does not bias the results.
Two choices have been tried, either a constant potential inside the
central part (that implies potential discontinuities at the boundaries)
or a linear form matching the value of the potential at electrodes.
The first one seems more
appropriate for a metallic situation in which the electric field inside
the sample vanishes whereas the second choice could describe an isolating
behavior that allows the formation of a large electric field inside
the central part. Notice that both choices imply the accumulation of
charge at the electrode boundaries. Results for the first choice will be
presented in the paper although checks have systematically been run
to assure that conclusions are model independent.

\begin{figure}
\includegraphics[width=0.45\textwidth]{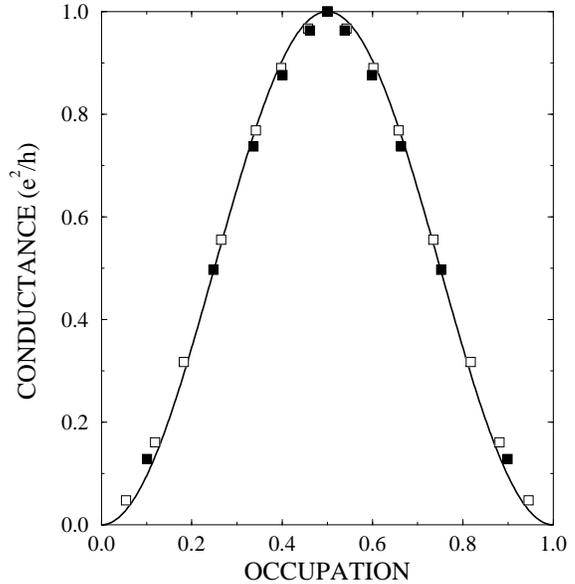}
\caption{Relation between the conductance through a single Anderson impurity
and its charge. Numerical results corresponding to $U=4t$ and several values
of the impurity level from $-5t$ to $t$ are given by squares 
(open for a central cluster made by seven sites and filled for the
larger cluster formed by eleven sites) and
compared with the exact result based on the Friedel--Langreth sum rule
(continuous line).}
\label{langreth.ps}
\end{figure}

The second issue has to do with the correct description of the interfaces
between electrodes and sample. Since our main approximation to the
many--body nature of the problem comes from the use of a selfenergy
obtained for a finite system, the inclusion of one or more non--interacting
sites within this finite part seems appropriate. Unfortunately,
the correct description of a wide band metal requires more than one
orbital per site and a precise model for the one--electron hopping
elements. Note that the use of a Hamiltonian like 
(\ref{hubbard}) with $U=0$ for the electrodes could produce important
deviations from correct results. For example, when electrons traverse
the interacting part through the upper Hubbard band at energies
around $U$, the electrode should have finite density of states at
this energy to allow electronic transport. This is not the case for
the chain model with a band between $-2 t$ and $+2 t$ when $U$
is large enough ($U>2t$ suffices).
The same happens at half--filling when a Kondo--like resonance
allows good electronic transmission at $U/2$.
In this work in which methodology primes over real world, we will
use electrodes described by a single band of $4 t$ width {\em but}
conveniently shifted to allow the electronic current coming from
the interacting part\cite{electrode}.
Therefore, when some part of the electrodes
is included in the heavy part of the calculation,
atomic levels will be $0$ when transport is through the lower
Hubbard band, $U$ when transport is through the upper Hubbard band
and, $U/2$ at half--filling when resonances within the gap could
eventually allow the charge transit. Non--interacting sites
included in the Lanczos part of the calculation
will be referred as interface region or salvage sites
in the rest of the paper.

\begin{figure}
\includegraphics[width=0.45\textwidth]{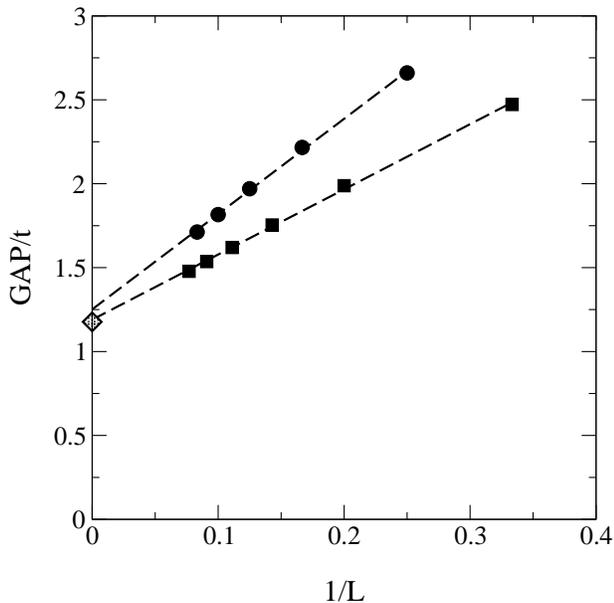}
\caption{Size dependence of the gap of the Hubbard
model at half--filling for $U=4 t$.
Circles (squares) give results for even (odd) number of sites values $L$
while the diamond gives the exact value of Lieb and Wu corresponding to
infinite $L$. Straight lines have been obtained by linear regression
to the numerical data.}
\label{thegap.ps}
\end{figure}

\begin{figure}
\includegraphics[width=0.45\textwidth]{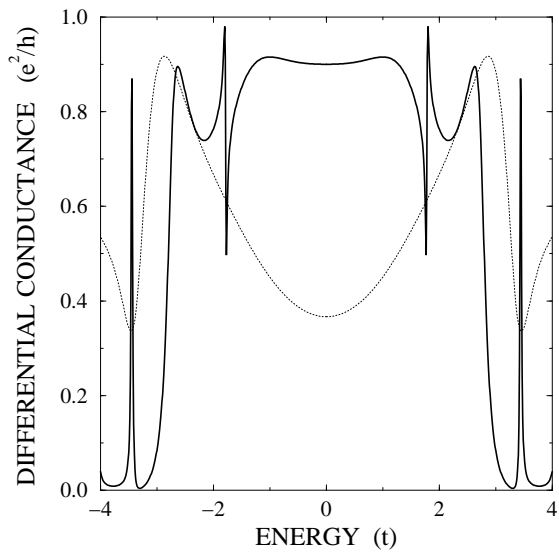}
\caption{Transport through three interacting sites directly
coupled to the electrodes (dotted line) or coupled to the electrodes
through one salvage site at each side (thick continuous line).
Results are given for $U=\pi t$.}
\label{ejemplo.ps}
\end{figure}

Now, we are prepared to list the numerical steps that are followed
to get the current-voltage characteristic of an arbitrary system.
First, the groundstate
for a given cluster occupation is obtained and, second, the tridiagonal
form of all needed Green functions is evaluated.
Third, the chemical potential of the finite cluster is obtained from
two additional groundstate Lanczos calculations for one more and
one less particle (the arithmetic mean is actually used).
Fourth, the external voltage
difference $V$ is divided in equal parts
that are added (subtracted) to the chemical potential (the one obtained
in the previous step) of the left (right) electrodes.
Five, Eq. (\ref{keldysh2}) is used to obtain the Keldysh Green function
that provides the non--equilibrium
description of the coupled system. With these ingredients,
application of Eq. (\ref{cur}) allows the calculation of the
current--voltage characteristic of the whole system.

Although straightforward and non--biased, the sketched numerical
recipe implies non--checked approximations. The results presented
in subsections B and C provide some credibility to the method
because they are physically meaningful and consistent.
Nevertheless, we have conducted a more specific test for just
a one site system: the one--dimensional Anderson impurity problem.

\subsection{Anderson impurity}

In this problem only electrons at one particular site labeled $0$
suffer interaction via a Hubbard term. Both the one--electron
energy of this particular level and its hopping to nearest--neighbors
are usually different from the rest:

\begin{eqnarray}
\lefteqn{
{\widehat {\cal H}_{imp}} =
\sum_{<i,j>,\sigma} t_{ij} {\hat c}^{\dag}_{i,\sigma} {\hat c}_{j,\sigma} +
}
\nonumber\\
& & {} \epsilon_0 \sum_{\sigma} {\hat n}_{0,\sigma}
+ U {\hat n}_{0,-\sigma} {\hat n}_{0,\sigma} \qquad ,
\label{kondoham}
\end{eqnarray}

\noindent
where $t_{ij}=t$ everywhere except at the impurity site. In effect,
$t_{01}=t_{10}=t_{0{\overline1}}=t_{{\overline1}0}=0.3t$
have been taken as representative of a weakly connected dot with
more or less relevant many--body interactions given by $U$.

Notice that Hamiltonian (\ref{kondoham}) describes an infinite chain
while Hamiltonian (\ref{hubbard}) in IIA refers to a finite cluster.
Our theoretical method is easily adapted to this problem considering
the interacting site plus some additional sites at each side as the finite
system while the remaining semiinfinite chains are now the electrodes.

Firstly, the system has been studied at half--filling, i.e.,
when just one electron occupies the central site.
This happens when the impurity level $\epsilon_0$ lies at $-U/2$.
One or more salvage sites have been added to each impurity side to
allow the coupling of the spin at the impurity with the spin of the
conduction band electrons. An odd number of non--interacting sites
should be added to each site of impurity to allow the coupling among
band states at its center and the impurity level at the same energy.
Since the hopping between impurity and lateral chains is usually much
smaller than $t$, even number of salvage sites giving states in
$\pm \epsilon$ pairs do not couple appreciably to the impurity and
are useless for our purpose. In this way, we typically end
with a system formed by
an odd number of sites with and odd number of electrons. The corresponding
groundstate is a doublet. We take care of the spin degeneracy of
the groundstate including a spin index in all Green function matrices
that consequently double their dimension.
Doing so, final results do not depend on the particular
linear combination chosen as the groundstate.

It is well--known that
a sharp resonance develops at the Fermi energy giving
rise to perfect transmission through the one site cluster (Kondo
resonant tunneling\cite{kondo}).
Figure \ref{kondo2.ps} shows this resonance and how it becomes
narrower as interaction $U$ increases.
Although differential conductance has been plotted as a function
of the conducting energy window $e V$, results for the spectral
weight on the impurity site are very similar. The results shown in the
figure have been obtained for a finite cluster formed by
eleven sites with the impurity at its center after connecting the
electrodes. While the $U$ dependence
of the resonance width follows the correct trend,
an exponentially narrowing of the resonance is not obtained.
Actually, the width of the resonance follows a $U^{-2}$ law
that is shown in the inset of the Figure.

Secondly, the system has been displaced from half--filling varying the
impurity level. The number of electrons in the cluster is fixed by the
chemical potential of the metallic band. Typically, we have one electron
on every system site except at the impurity which occupation per spin
varies from 0 to 1 as its level moves from $+ \infty$ to $- \infty$.
Transmission at the Fermi level is close to 1 when occupation is close
to $0.5$ and
decreases rapidly when the number of electrons on the impurity site
changes. The precise relationship between charge and conductance is
determined by the Friedel--Langreth sum rule (see, for example,
Eq. (7) of Ref.(\onlinecite{kondo}) and Ref.(\onlinecite{langreth}).
When the impurity is symmetrically coupled to both semiinfinite
chains, the relation is given by:

\begin{equation}
G_{\sigma}={\sin}^2 (\pi <n_{0,\sigma}>) \qquad .
\label{friedellangreth}
\end{equation}

Figure \ref{langreth.ps} gives results
for clusters formed by seven and eleven sites connected to half--filled
electrodes. Here, the value of $U$ is always $4t$ while the impurity level
takes values between $-5t$ and $t$.
It can be seen that the fulfillment of the exact relation
is excellent regardless the number of sites used to obtain the many--body
selfenergy\cite{sutileza}.

We conclude that the Anderson impurity is well described by our numerical
procedure even though a naive estimation of the extension of correlations
suggest many more sites than included in the exact cluster calculation.
Notice that although correlation selfenergy is local, it produces
long distance effects well within the semiinfinite chains. 
Pictorially, one can say that any state coming from one electrode
suffers scattering by the many--body selfenergy
at the impurity site when traveling to the other
electrode. In our scheme, correlation effects among these extended states
just come from the components
that they have on the finite cluster states that are exactly
described by the Lanczos computation.

\subsection{Half--filled system}

Half--filling of the Hubbard model happens when
the number of electrons equals the number of sites.
Half--filled Hubbard model is an isolator because charge excitations
occur through an energy gap which is of the order of $U$ for large
interaction values\cite{lieb,drude}.
This exact result proved by Lieb and Wu using the Bethe ansatz is
easily recovered by a finite cluster calculation both for even and
odd number of sites (and a corresponding equal number of electrons)
in spite of the apparent difficulty of dealing with an odd number
of electrons in half the cases. When the number of sites is odd, either
an up spin electron or a down spin electron is unpaired but for any
choice the total density of states shows a gap. Partial up spin and
down spin densities of states are different but share the same gap feature.
Integration of these densities of states up to the gap gives the
number of up and down spin electrons really present in the system.
Fig.(\ref{molina2.eps}) shows an example density of states for eleven
interacting sites while Fig.(\ref{thegap.ps})
shows how the gap scaling toward the exact Lieb and Wu result proceeds.
Results for clusters formed by even or odd number of sites and electrons
scale along different straight lines that converge to the same limit.
The gap is systematically smaller for the spin degenerate case. If one finds
a spin polarized groundstate (and consequently, a spin polarized
density of states) uncomfortable, a mixed groundstate can be formed
as a linear combination of $S_z={1 \over 2}$ and of $S_z=-{1 \over 2}$
states and spin polarization disappears when the weight of both components
equals. The gap in the density of states does not change.
This behavior has a very simple physical explanation that has very
important consequences for transport properties. No matter the spin
of electrons, the energy cost for one less or one more electron
within the system is about $U$ at half--filling because every site is
occupied by just one electron. Transport of electrons through the
system implies a change of 1 in the number of electrons of the system
and consequently an energy cost of the order of $U$. Electrons having
just a bit more energy than the chemical potential in a d.c. experiment
are unable to traverse the cluster. More precisely, the Green function
of the coupled system develops exponential tails within the Hubbard
cluster and conductance should scale toward zero for large
enough systems. Our results below give a detailed description
of this situation.

\begin{figure}
\includegraphics[width=0.45\textwidth]{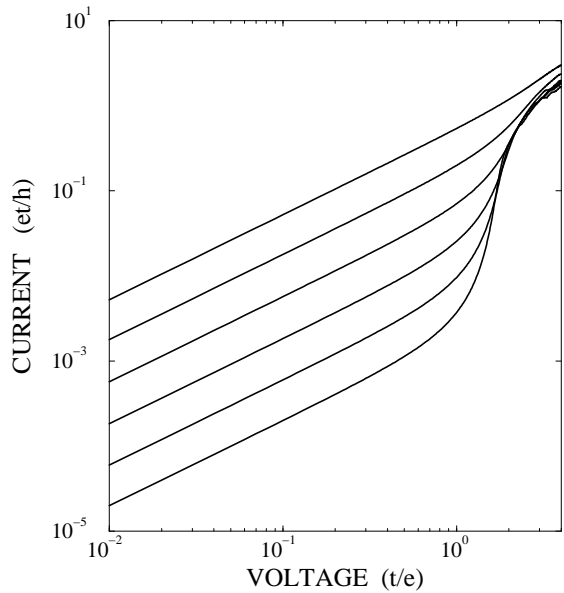}
\caption{This figure illustrates the size scaling of our results at
half--filling for an arbitrary value of the interaction $U=4t$.
Curves correspond to system lengths $L$ from 2 (top) to 12 (bottom) in
steps of two sites. The exponential dependence on chain length
is manifested by the parallelism of current--voltage curves
at voltages below the charge gap. The same trend is followed at
other $U$ values although the exponential scaling of the current
begins at larger lengths for smaller interactions.}
\label{scalingL.ps}
\end{figure}

Previous paragraph has shown that odd clusters have nothing special
in spite of spin degeneracy. Therefore, it is quite surprising that
Oguri\cite{oguri,oguri2}  claims perfect transmission through
interacting clusters formed by an odd number of sites larger than one
based -at least partially- in spin degeneracy. Notice that the only way
of getting perfect transmission is the presence of a state (or resonance)
at the chemical potential, i.e., at the gap center in the symmetric case.
While this state is absent in an exact cluster calculation, it is
present in Oguri's Green functions because an odd number of one--electron
states implies a state at the bandcenter and the exact but just
second--order perturbation theory in the electron repulsion $U$
that he performs is unable to open a gap.
Precisely, this was the reason given by
Oguri for his result in the first paper of the series\cite{puntilla}.
What is then the physical reason for perfect transmission across
{\it one} interacting site (Anderson resonance) but not for three or more
interacting sites? Although not a mathematical argument there is a
simple reason for this behavior: states in the whole system can be
classified as even or odd according to their reflection symmetry
relative to the central atom of the interacting region. When this
region is formed by a single atom {\em but only} in this particular case,
odd states do not couple to the interaction. Then a Fermi picture
for the whole interacting groundstate is possible in which
some of the occupied states are purely one--electron
states. If the last occupied state corresponds to one of these
states, perfect transmission appears. This is always the case for an
extended system but just happens in our finite cluster calculations when
the number of odd states is odd and one of these purely one--electron
states defines the Fermi level. This means 1+1+1
(the first integer gives the
number of sites at the left side of the interacting region, the second one is
the number of interacting sites and, the last integer the number of sites at
the right side of the interacting region), 3+1+3, ... systems but
not 2+1+2, 4+1+4, as we have seen in section IIIA. As a direct consequence
of this naive picture of the real complex many--body groundstate,
it is clear that the number of electrons that
occupy the non--interacting state at the chemical potential
can be 0, 1 or 2 without energy change.
We have checked that this is indeed the case for the clusters used
in IIIA (the actual multiplicity of the cluster groundstate is four:
one spin doublet for which number of electrons and sites coincide,
a spin singlet with one extra electron and a second
spin singlet with one less electron). This situation directly leads
to a Green function showing a zero energy {\it charge}
excitation that explains the appearance of perfect transmission.
Summarizing our point of view,
there is a nice physical reason for perfect transmission
across one interacting site but the persistence of perfect transmission
across large interacting regions contradicts the insulating character
proved for the Hubbard model by Lieb and Wu\cite{lieb}.

Let us come back to a detailed numerical
study of transport through half--filled
systems once the overall incorrectness of Oguri results has been shown.
Based on familiar results of a standard one--electron isolating material,
one can naively think that current--voltage characteristic
will show an activated
behavior for voltage values above the charge excitations gap.
In fact, this is the result if the selfenergy corresponding to a pure
finite Hubbard chain is used in the transport equation (\ref{cur}).
It should be so since the many--body selfenergy entering into the
global Green function describes the charge gap.
This behavior does not change if one or more salvage sites are
added to each side of the cluster that is numerically solved.
Nevertheless, scaling proceeds differently for clusters formed
by an even number of sites (and electrons) which groundstate is
a singlet and clusters of odd numbers of sites (and electrons) which
groundstate is a doublet. In the last case, we can speak of Kondo--like
systems that show a tendency to form resonances with good transmission
properties at the chemical potential energy $U/2$.
Just to give a first example, Figure \ref{ejemplo.ps} shows the
qualitative change in the transport properties of Kondo--like clusters
when interface sites are or are not included in the selfenergy calculation:
while the conductance of the system that completely ignores the electrodes
has a minimum at the chemical potential due to the formation of a
pseudogap, transmission is almost maximum over a large energy range
when one electrode site is included at each side. Physically, this
result can be interpreted saying
that the unpaired electron of the small cluster forms a singlet
with the band state at the chemical potential {\it only} if some
electrode sites are included in the calculation of the correlation
selfenergy. We will begin showing the behavior of the simpler
systems (even number of sites) and come back to the degenerate case
(odd number of sites) at the end of the subsection.

\begin{figure}
\includegraphics[width=0.45\textwidth]{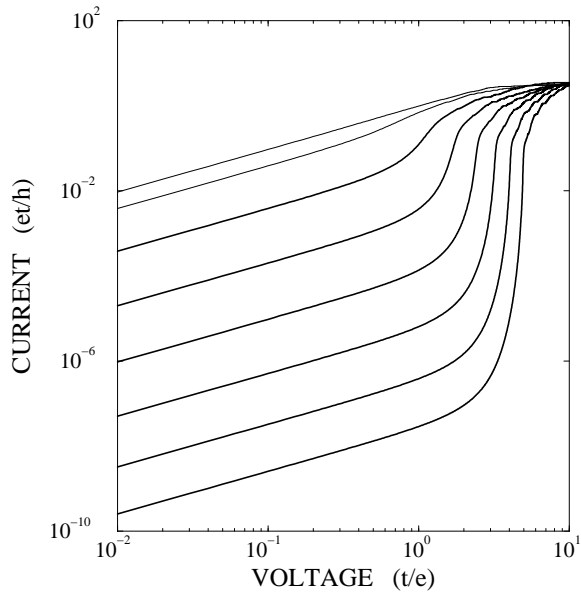}
\caption{This figure illustrates the $U$ dependence of the current--voltage
curves at half--filling for a system of fixed size (twelve sites).
$U$ goes from $t$ (upper line) to $8t$ (lower line) in $t$ steps.
As in the previous figure, the exponential dependence
is manifested by the parallelism of current--voltage curves
at voltages below the charge gap. This happens for an $U$ value larger
than $3t$ for this system size.
The systematic gap enlargement due to the increasing value of $U$
is also shown by the Figure.}
\label{scalingU.ps}
\end{figure}

Size scaling of the problem at hand is shown in
Figure \ref{scalingL.ps}.
Current--voltage curves have been obtained for Hubbard chains
of increasing length (from two to twelve sites) and interaction $U=4t$.
The number of electrons in the chain equals the number of sites.
The chemical potential is exactly $U/2$ in this situation.
Figure \ref{scalingL.ps} shows an exponential decrease of the current
as a function of the length of the interacting part:
$$
I \propto \exp (-\alpha L) \qquad ,
$$
which implies
$$
\log (I) \propto - \alpha L \qquad .
$$
The linear $L$ dependence in the logarithmic scale can be observed
for any value of the potential below the gap
(which is about $1.75 t$ for $U=4t$).
This exponential scaling law is valid for any value of $U$.
Consequently, a large enough interacting region completely prohibits
current below the gap energy at half--filling. 

\begin{figure}
\includegraphics[width=0.45\textwidth]{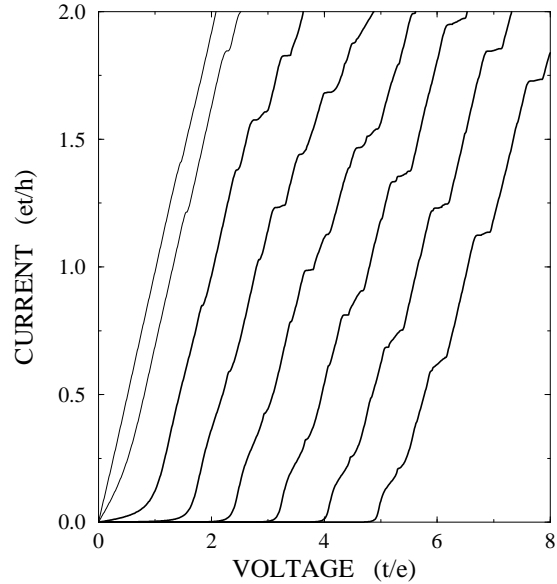}
\caption{This figure illustrates the $U$ dependence of the gap
at half--filling for a system of fixed size (twelve sites).
The data of previous Figure are plotted here in linear--linear scale:
$U$ increases from $t$ to $8t$ in $t$ steps for curves from left
to right. Current--voltage characteristics show here
in a more clear fashion the activated behavior of current
for voltage values above the gap charge.}
\label{gapU.ps}
\end{figure}

Figure \ref{scalingU.ps} shows the $U$ dependence of a system
formed by twelve sites. From top to bottom the IV curves
for $U/t$ values from 1 to 8 (in steps of one unity) have been plotted.
It can be seen that the $U$ dependence is quite similar to the
previously seen $L$ dependence. As before, current exponentially
vanishes as a function of $U$ (for $U$ values larger than $3t$ at this
size) for any potential value below the charge gap.
The same data have been plotted in Figure \ref{gapU.ps} in
a linear--linear scale to emphasize the behavior of the current
for energies above the gap. It can be seen that an almost linear increase
of almost maximum conductance slope takes place regardless the value
of $U$. That means that electrons can enter the interacting system at
energies in the lower Hubbard band and exit at energies corresponding
to the upper Hubbard band suffering just a small amount of
scattering. We will discuss this apparent absence of scattering once
the results at fillings different from half (metallic situation)
are given.

\begin{figure}
\includegraphics[width=0.45\textwidth]{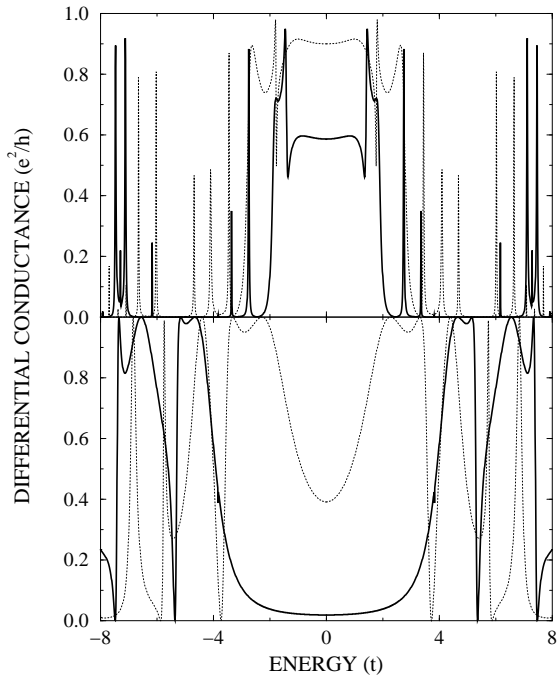}
\caption{Differential conductance of small clusters for $U=\pi t$
(thin dotted lines) and $U=2 \pi t$ (thick continuous lines).
Upper panel: three interacting sites embedded in a five sites cluster.
Lower panel: four interacting sites.}
\label{oguri_bis2.ps}
\end{figure}

Let us do a detailed comparison with Oguri results for the
electronic transport through three and four interacting sites connected to
reservoirs\cite{oguri2}. Our results are shown in Fig. \ref{oguri_bis2.ps}
and should be compared with panel (a) (corresponding to zero temperature)
of Figures 10 and 11 of the mentioned paper\cite{pega}.
While transmission across the three--sites system are comparable,
in particular for energies within the non--interacting band,
results for the four--sites system have nothing in common.
Our results show the appearance of a pseudogap for the smaller $U$
value and a well-developed gap for $U=2\pi t$
(larger than the one--electron bandwidth)
while Oguri results for this size are similar to the
ones for the smaller system. This result is a direct manifestation
of the impossibility of forming a gap using the perturbative method of
Ref. (\onlinecite{oguri2}). 
A closer look of the three--site system shows
that our calculated transmission at the bandcenter is not 1 (as has been
discussed before) although values close to 1 are reached
at other energies.

\begin{figure}
\includegraphics[width=0.45\textwidth]{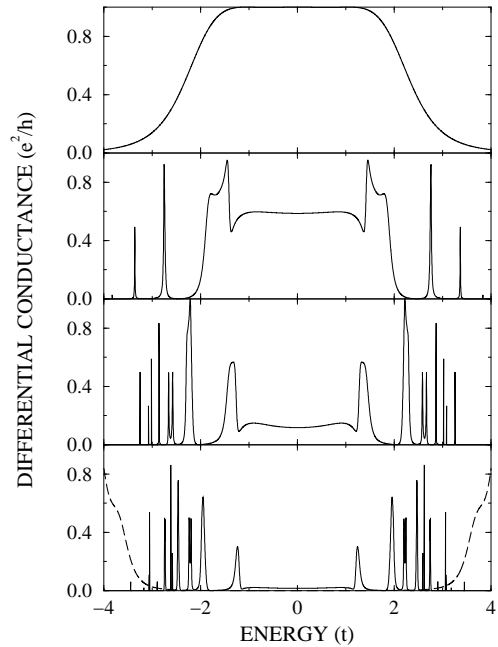}
\caption{Differential conductance of clusters formed by an odd
number of sites for $U=2 \pi t$. From top to bottom: One, three, five
and seven interacting sites plus one salvage site at each cluster side.
The dashed line in the bottom panel gives the differential conductance
of a seven sites cluster with no salvage sites: the presence of a gap
is evident.}
\label{summum.ps}
\end{figure}

\begin{figure}
\includegraphics[width=0.45\textwidth]{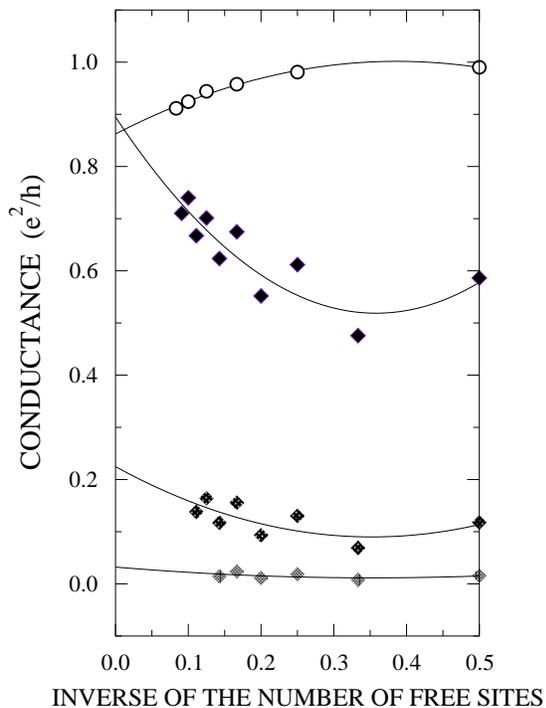}
\caption{Dependence of the conductance at zero bias on the number
of salvage sites included in the Lanczos finite cluster calculation.
Circles and diamonds of decreasing density give results for two, three,
five, and seven interacting sites when $U=2 \pi t$.
Continuous lines give parabolic extrapolations to the
infinite limit. Configurations with an even number of sites at both
sides have not been considered as they artificially drop conductance.
For example, eight salvage sites are asymmetrically distributed:
three at one side and five at the other.}
\label{chiappe.ps}
\end{figure}

The size dependence followed by spin--degenerate clusters is exemplified
in Figure \ref{summum.ps}. From top to bottom the
size of the interacting part is increased from one site to seven sites.
It is evident that transmission rapidly diminishes with system size:
in effect, the result at the bottom shows narrow resonances in a
developing gap (dashed line gives the result excluding electrode
sites, i.e., avoiding the formation of singlet cluster--band states).
Figure \ref{chiappe.ps} shows the dependence on the
number of salvage sites included in the Lanczos finite cluster
calculation. Conductance at zero bias is given for two, three, five
and seven interacting sites for cluster sizes up to fourteen sites.
While the addition of salvage sites somewhat diminishes the
conductance through two interacting sites, it increases
the conductance of systems formed by an odd number of sites.
Numerical data have been fitted by parabolas both for simplicity
and because an $L^{-1}$ fit has been quite successful describing the
gap formation (see Fig.(\ref{thegap.ps}).
Although an extrapolated value of 1 cannot be excluded for the
three sites result, it seems quite difficult that the data for five
and seven sites (two lower parabolas) extrapolate to 1.

\begin{figure}
\includegraphics[width=0.45\textwidth]{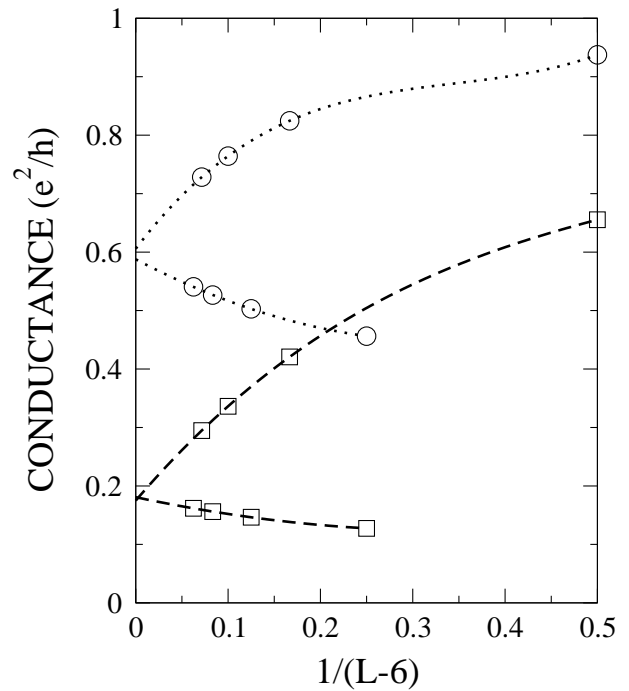}
\caption{Scaling of the conductance of a connected chain of
six sites in which spinless fermions interact according to the
model of References \onlinecite{spinless1} and \onlinecite{spinless2}
as a function of the inverse of the number of salvage sites.
Circles correspond to $U=2$ ans squares to $U=4$. Polynomial fits take care
of the even/odd dependence in the number of salvage sites at each side.
Extrapolation gives conductances $0.60 \pm 0.01$ and
$0.177 \pm 0.003$ for $U$ equal to 2 and 4, respectively.}
\label{fit.eps}
\end{figure}

In conclusion, our results show that when size scaling is correctly
taken into account, all systems converge to the same physical behavior
at half--filling: current diminishes exponentially at voltages
below the charge excitations gap just due to the existence of this gap.
Larger $U$ values accelerate the scaling. At the end, no odd--even
differences survive. Nevertheless, it is clear that the correct
treatment/scaling of Kondo--like systems (formed by any odd number of
impurities) requires much more care
since results for small systems strongly depend on the details (presence
or not of salvage sites, the precise description of these {\it
metallic} sites, the number of them, etc.).

At this point, a comparison of our results with recently published
work studying persistent currents in rings\cite{spinless1,spinless2}
is in order. In these works an interacting model of
spinless fermions is studied instead of the Hubbard model to
keep computational needs at a minimum level and make possible
a clear extrapolation
to the case of an infinitely large non--interacting part. 
Based on a result that has been proved for non--interacting systems,
the conductance through a finite interacting region can be
obtained from the comparison of the ring currents currents before and
after switching on interaction.
Here, we will compare the conductances presented in
Fig. 2 of Ref. (\onlinecite{spinless1})
and Fig. 3 of Ref. (\onlinecite{spinless2})
with the results obtained in our scheme when the Hubbard interaction
between electrons of opposite spins at the same site
is substituted by the following model of
spinless fermions interacting with particles at
nearest--neighbors sites:

\begin{equation}
U \sum_{i=1}^{{L_s}-1}({{\hat n}_i}-1/2) ({{\hat n}_{i+1}}-1/2) \qquad .
\label{spinless}
\end{equation}

Let us begin discussing the more conventional case of an even number of sites
$L_s$ occupied by ${L_s}/2$ fermions. We take $L_s=6$ (fermions on
these sites interact with other fermions at nearest--neighbors sites)
and add from one to eight salvage sites at
each side of the interacting part (fermions just hop over these sites but
no additional interaction exists). First, exact Green functions are calculated
for this finite cluster following our standard procedure and then
Eq. (\ref{keldysh2}) is used to get approximate Green functions of the
infinite connected system. Linear response conductances are
obtained from non--diagonal Keldysh Green functions at the chemical
potential which is zero at half--filling (see Eq. (\ref{dc})).
Figure \ref{fit.eps} shows our results
for $U$ values 2 (circles) and 4 (squares) together with polynomial
extrapolations to the limit of a large number of salvage sites. Conductance
strongly depends on the even/odd character of the number of extra
sites added to each side of the chain.
Therefore, regression should take care of this
particularity but converges to a more or less well--defined value. Our
extrapolated conductances nicely agree with the values obtained for rings for
both $U$ values. Actually, the oscillatory convergence of the
conductance to the infinite size limit gives additional credibility
to the existence of a well--defined finite value of the conductance.

Turning now to the case of an odd number of sites $L_s$ and repeating the
procedure a first difficulty appears: $L_s/2$ is not an integer number but
an integer number of particles is needed to proceed. Fortunately,
groundstate energies for occupations $(L_s+1)/2$ and $(L_s-1)/2$ coincide
and a linear combination of both degenerate states
with equal weight $1/2$ produces
a Green function describing perfect half--filling, that is,
showing half fermion per site.
Using the same linear combination, exact Green functions are obtained
for larger clusters that include additional non--interacting sites.
Showing a behavior quite dissimilar to the one found for the Hubbard model,
conductance of any of these systems is exactly 1 when connected to
leads via the Dyson equation. Therefore, conductances for spinless
fermions interacting on an odd number of sites perfectly
agree with the findings of works (\onlinecite{spinless1}) and
(\onlinecite{spinless2}). But the initially surprising results have an easy
interpretation in view of the strong differences shown by the
charge excitation spectra of both model. Figure \ref{molina2.eps} compares
exact density of states of eleven site clusters described either
by the Hubbard model (upper panel) or by the spinless interacting
model (lower panel). In both cases $U$ is equal to 4 and a large imaginary
part of $0.1 t$ has been added to the excitation energy to show the
$\delta$--functions of the spectrum. Both are perfectly symmetric
around 0 but while the Hubbard model shows a clear gap around the
chemical potential, the spinless fermions model has a state at this energy.
Moreover, even the existence of a well--defined gap is not so clear
for this model\cite{futuro}. The zero energy excitation comes from the
ambiguous number of particles inside the system, i.e., one more or one
less particle comes into the system without an energy change.
This particular excitation is the responsible -when coupled to
the leads- of the perfect transmission at the chemical potential energy.
On the other hand, 
the Hubbard model shows just a particle per site at half--filling and
a typical cost for adding or removing one electron is $U/2$.
As discussed above, although a dc current can traverse {\it small}
Hubbard systems, their value systematically diminishes with the
size of the system. In short,
the particular behavior of the spinless interacting system comes precisely
by the substitution of the spin degeneration of the Hubbard model on systems
of an odd number of sites by a charge degeneration that allows
charge transport at zero energy cost, i.e., finite -actually perfect-
dc conductance.

In this way, our whole numerical procedure has been indirectly
checked through the
comparison to the results obtained by the
alternative numerical methods used to get conductances
of spinless fermions through small interacting regions.

\begin{figure}
\includegraphics[width=0.45\textwidth]{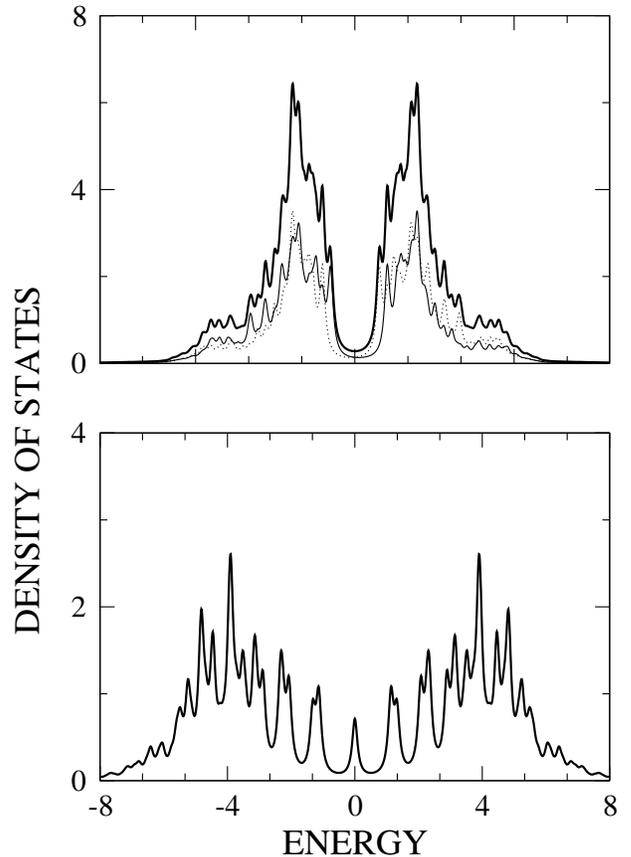}
\caption{The density of states of an isolated Hubbard chain of
eleven sites at half--filling (six electrons with spin up and five
electrons with spin down) (upper panel) is compared with the same
magnitude for the spinless fermionic model of
References (\onlinecite{spinless1}) and (\onlinecite{spinless2})
on a chain formed by eleven sites containing 5.5 fermions (lower panel).
In both cases the value of $U$ is 4 and the imaginary part added
to the energy to show $\delta$--functions is 0.1.
Partial densities of up (continuous line) and down (dotted line) spin
states are also given. Energies are given in terms of the
hopping energy $t$ while $t^{-1}$ is used
for the density of states.}
\label{molina2.eps}
\end{figure}

\begin{figure}
\includegraphics[width=0.45\textwidth]{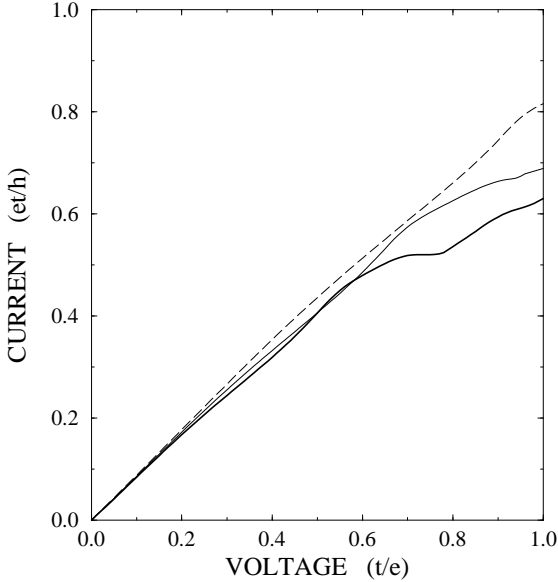}
\caption{This figure illustrates the size independence of the
metallic behavior of the interacting system at one--quarter--filling.
Current--voltage characteristics for lengths eight (dashed line),
twelve (continuous line) and sixteen (thick line) are compared.}
\label{quarterL.ps}
\end{figure}

We would like to close this subsection concluding that
while our results support the accepted idea that Hubbard model
is an insulator at half--filling\cite{drude} they also give further
insight onto the transport of electrons through an interacting region.
For example, the large value of the differential conductance for
energies above the gap and its independence on the $U$ value comes
somewhat as a surprise. While small scattering is quite plausible for
small values of $U$, one could expect large scattering for $U$ values as
large as twice the one--electron bandwidth.

\subsection{Metallic system}

Results for systems away from half--filling are presented in this
subsection. In this situation, chemical potential is no longer $U/2$ as
before and should be obtained for the finite
cluster from groundstate calculations with different number of
particles. We define:

$$
\mu_+ = E_0 (N+1) - E_0 (N)
$$
$$
\mu_- = E_0 (N) - E_0 (N-1) \qquad,
$$

\noindent
where $E_0 (N)$ represents the groundstate energy corresponding
to $N$ particles and take the arithmetic mean of both:

$$
\mu = (\mu_+ + \mu_-)/2  \qquad.
$$

The difference between both estimations is typically very small in this
case (actually both should coincide away from half--filling in a closed
Hubbard chain\cite{lieb}). We always have $\mu < U/2$
when the number of particles is smaller than the number of sites.
As said at the beginning of this section,
this value of the chemical potential is also assumed for the connected
system. Numerical result show that this assumption is internally
consistent: total number of electrons in the cluster and their spatial
distribution hardly depend on the electrodes presence.

\begin{figure}
\includegraphics[width=0.45\textwidth]{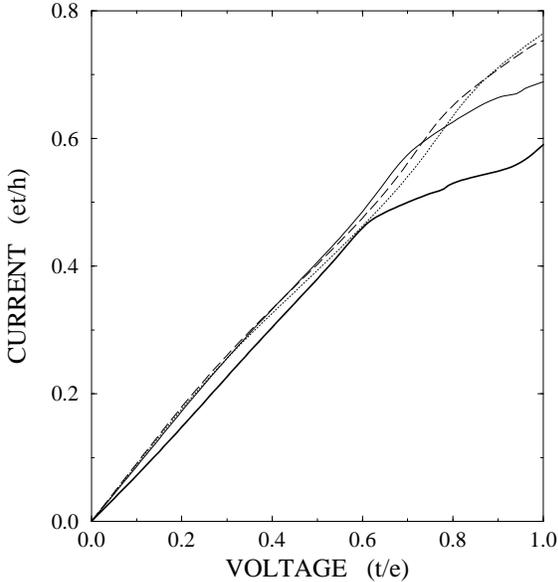}
\caption{This figure illustrates the $U$ dependence of the current--voltage
curves in the metallic regime at one--quarter--filling.
The correlation selfenergy has been obtained for twelve sites
and several $U$ values.
Results for $U/t=1,2,4$ and, $8$ are given by dotted, long--dashed,
thin and thick curves, respectively.}
\label{quarterU.ps}
\end{figure}

The more important result in the metallic case is the length independence
of the transport through the interacting part. This is shown by
Fig. \ref{quarterL.ps} for the current through eight, twelve and
sixteen Hubbard sites at one--quarter filling.
For this comparison, a relatively large value $U=4t$ of
the interaction has been chosen. Differences between results corresponding
to different lengths are probably measuring the typical precision
of our approach. Additional numerical calculations not discussed in
this paper help to show that all scattering occurs
at the system contacts but not inside the interacting region.
Let us proceed to the evaluation of the contact resistance and its
dependence on $U$ and filling factor.

\begin{figure}
\includegraphics[width=0.45\textwidth]{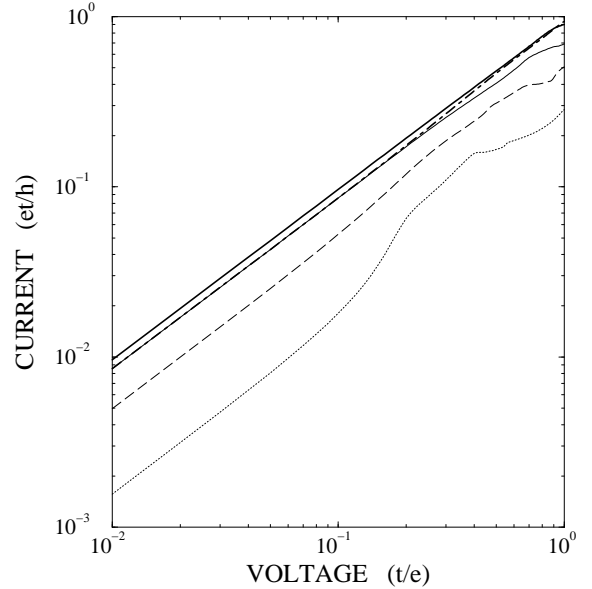}
\caption{This figure illustrates the metallic behavior of the interacting
system at fillings others than half.
Results correspond to twelve sites, $U=4t$ and increasing number of
electrons: $Q=2$ (dotted curve), 4 (long--dashed curve),
6 (thin continuous line), 8 (dot--dashed curve), and 10 (thick
continuous curve). Notice that the last three curves are very close to
maximum differential conductance (slope 1 in the units of the figure).}
\label{quarterQ.ps}
\end{figure}

Figure \ref{quarterU.ps} gives current--voltage curves for a chain
of twelve interacting sites directly connected to metallic
electrodes of constant (and large) density of states.
The number of electrons within the cluster is always six while
$U$ increases from $t$ to $8t$. Scattering at the contacts is always
small regardless the large values of $U$ included in the Figure.
For example, the result for $U=8t$ (twice the corresponding
non--interacting bandwidth value) at $V=0.5 t/e$
can be explained by a transmission coefficient about
0.87 at each interface.
Actually, the approximate independence on $U$ is quite surprising.

Figure \ref{quarterQ.ps} gives current--voltage curves for the
same system but varying the number of electrons in the cluster
from 2 to 12 in steps of 2. In all cases a linear characteristic
is obtained (better for small voltages) but the slope of the curve,
i.e., the differential conductance,
{\em strongly} depends on the filling factor (note that slope
of the IV curve is given by the current value at the smaller voltage
value in a log--log graph).
While differential conductance is close to its maximum value of one
quantum for large number of electrons (eight or ten), it is much
smaller for small filling factors (two to six electrons in twelve
sites). More precisely, average conductance is $0.29$ for two electrons,
$0.51$ for four electrons and $0.69$ for six electrons (one--quarter
filling factor). Although it is tempting relating these values to
experimentally measured values of conductances in clean quantum wires,
we think that an improved geometrical model is necessary prior to a
realistic comparison with experiments (For example, the presence of
several conducting channels that can be active as a function of the Fermi
level should be accounted for). Moreover, the consideration of long--range
interactions among electrons can also change conclusions.
In spite of these difficulties for approaching the real world
that will be explored in further work,
it seems interesting
that the major (may be the unique) dependence of transport across
an interacting system comes from the filling factor (from the
chemical potential as observed from the electrodes) and not from the
value of the interaction.

\section{Summary}

A general and computationally convenient method has been implemented
for the evaluation of the current--voltage characteristic through
interacting regions. The method has been checked studying an isolated
Anderson impurity in a linear chain. Two main results have been proved
for extended Hubbard systems. First, the isolating behavior at
half--filling including the possibility of activated transport
for voltage values above the gap.
Our results confirm the physical picture obtained studying the
scaling of the Drude peak of half--filled Hubbard rings\cite{drude}
but disagree in important details from previous studies in the same
open geometry\cite{oguri,oguri2}. For example, Kondo--like systems
formed by an odd number of impurities show a resonance transmission
that decays exponentially with the number of sites.

Second, good metallic behavior has been obtained away from
half--filling when the number of electrons is large enough.
In this case,
the interacting region allows ballistic electrons that are
just scattered by the contacts between electrodes and the Hubbard part.
The value of these contact resistances hardly depends on $U$ but
shows an important dependence on the filling factor or the
chemical potential from a different point of view.
Ballistic transport, that is, size independent
current for a given voltage, seems to exclude real scattering within
the interacting sample and, therefore, exotic consequences
of interaction at $T=0$. Although
previously reported theoretical estimations of
quasi--perfect transmittance\cite{maslov,pono,schulz}
are partially confirmed in this one--dimensional
metal--Luttinger liquid--metal setup just below half--filling,
conductance is smaller at other filling factors.
We guess that broader systems allowing several conductance channels
are needed to correctly model the experimental phenomena observed
in quantum wires.

\begin{acknowledgments}
E. V. Anda is acknowledged for fruitful discussions on the method.
One of the authors (JAV) thanks R. Aguado for helping
in the development of the Anderson impurity subsection.
This work has been partially supported by
Buenos Aires University (grant UBACYT-EX210),
Argentinian CONICET, Spanish 
Comisi\'on Interministerial de Ciencia y Tecnolog\'{\i}a (grants PB96-0085
and  MAT2002-04429)
and Direcci\'on General de Ense\~nanza Superior e Investigaci\'on y Ciencia
(grant 1FD97-1358).
\end{acknowledgments}


\begin{thebibliography}{}

\bibitem{maslov}
D. L. Maslov and M. Stone, Phys. Rev. B {\bf 52}, R5539 (1995);
D. L. Maslov, Phys. Rev. B {\bf 52}, R14368 (1995).

\bibitem{pono}
V. V. Ponomarenko, Phys. Rev. B {\bf 52}, R8666 (1995).

\bibitem{schulz}
H. J. Schulz, cond-mat/9503150,
{\it Proceedings of Les Houches Summer School LXI},
ed. E. Akkermans, G. Montambaux, J. Pichard, et J. Zinn-Justin
(Elsevier, Amsterdam, 1995), p. 533;
I. Safi and H. J. Schulz, Phys. Rev. B {\bf 52}, R17040 (1995).

\bibitem{oguri}
A. Oguri, Phys. Rev. B {\bf 59}, 12240 (1999);
{\it ibid.} {\bf 63}, 115305 (2001);
{\it ibid.} {\bf 64}, 153305 (2001).

\bibitem{oguri2}
A. Oguri, J. Phys. Soc. Jpn {\bf 70}, 2666 (2001).

\bibitem{ymas}
A. A. Odintsov, Y. Tokura, and S. Tarucha,
Phys. Rev. B {\bf 56}, R12729 (1997);
V. V. Ponomarenko and N. Nagaosa, Phys. Rev. Lett. {\bf 81}, 2304 (1998)
and {\it ibid.} {\bf 83}, 1822 (1999).

\bibitem{caroli}
C. Caroli, R. Combescot, P. Nozieres, and D. Saint--James,
J. Phys. C: Solid St. Phys. {\bf 4}, 916 (1971).

\bibitem{meir}
Y. Meir and N. S. Wingreen, Phys. Rev. Lett. {\bf 68}, 2512 (1992).

\bibitem{dagotto}
E. Dagotto, Rev. Mod. Phys. {\bf 66}, 763 (1994).
Non--diagonal elements of the Green function are obtained using
linear combinations of fermion operators. For example:
$$
<({\hat c}_i+{\hat c}_j)^{\dag} ({\hat c}_i+{\hat c}_j)>=
$$
$$
<{\hat c}_i^{\dag} {\hat c}_i>+
<{\hat c}_i^{\dag} {\hat c}_j>+
<{\hat c}_j^{\dag} {\hat c}_i>+
<{\hat c}_j^{\dag} {\hat c}_j>
$$

\bibitem{keldysh}
L. V. Keldysh, Zh. Eksp. Teor. Fiz. {\bf 47}, 1515 (1964)
[Sov. Phys.--JETP {\bf 20}, 1018 (1965)].

\bibitem{landau}
L. D. Landau and E. M. Lifshitz, {\it Course of Theoretical Physics},
Vol. 10 (Pergamon Press, Oxford, 1981).

\bibitem{mahan}
G. D. Mahan, {\it Many-Particle Physics}, Second Edition
(Plenum Press, New York, 1990).

\bibitem{aclaracion}
Although notation of Ref.(\onlinecite{landau}) is used in this paper,
Green function matrices have been arranged according to Mahan
prescription (see Ref.(\onlinecite{mahan})). This form
allows non--diagonal selfenergies having the
same properties as the corresponding Green functions.

\bibitem{c5}
A. L. Yeyati, A. Mart\'{\i}n--Rodero, and F. Flores,
Phys. Rev. Lett. {\bf 71}, 2991 (1993).

\bibitem{brasil}
M. A. Davidovich, E. V. Anda, J. R. Iglesias, and G. Chiappe,
Phys. Rev. B {\bf 55}, R7335 (1997);
V. Ferrari, G. Chiappe, E. V. Anda, and M. A. Davidovich,
Phys. Rev. Lett. {\bf 82}, 5088 (1999);
C. A. B\"usser, E. V. Anda, A. L. Lima, M. A. Davidovich, and G. Chiappe,
Phys. Rev. B {\bf 62}, 9907 (2000).

\bibitem{seminal}
S. Robert and K. W. H. Stevens,
J. Phys. C: Solid State Phys. {\bf 13}, 5941 (1980),
E. V. Anda, J. Phys. C: Solid State Phys. {\bf 14}, L1037 (1981) and,
A. S. da Rosa Sim\~oes, J. R. Iglesias, A. Rojo and B. R. Alascio,
J. Phys. C: Solid State Phys. {\bf 21}, 1941 (1988).

\bibitem{c5bis}
The role played by the electrochemical potential across a constriction
has been addressed, for example,  by
P. L. Pernas, A. Mart\'{\i}n-Rodero, and F. Flores,
Phys. Rev. B 41, 8553 (1990) and
A. Levy Yeyati, F. Flores and E. V. Anda,
Phys. Rev. B 47, 10543 (1993).

\bibitem{electrode}
Notice that a detailed description of levels and hoppings is only
necessary for the sites coming into the numerical finite cluster calculation.
The far part of electrodes is still described by a constant density of
states through the local Green functions given in (\ref{leadocc}) and
(\ref{leadvac}) for the left electrode and similarly for the right one.

\bibitem{kondo}
T. K. Ng and P. A. Lee, Phys. Rev. Lett. {\bf 61}, 1768 (1988).

\bibitem{langreth}
D. C. Langreth, Phys. Rev. {\bf 150}, 516 (1966).

\bibitem{sutileza}
When calculating the many--body selfenergy,
the number of electrons present in the small cluster equals the number
of sites. When the extended system is solved, a new Fermi level is defined
at the energy at which the number of electrons in the cluster remains
unchanged. The shift of the effective Fermi level diminishes for larger
clusters and would eventually be zero for large enough systems.
This procedure is consistent with the idea that electrons flowing
into or from the impurity come from very large distances.

\bibitem{lieb}
E. H. Lieb and F. Y. Wu, Phys. Rev. Lett. {\bf 20}, 1445 (1968).

\bibitem{drude}
Numerical calculations of the optical conductivity of the Hubbard model
in a ring geometry show an exponential decay of the Drude weight
as a function of the number of sites. This scaling indicates
an undoubtedly isolating behavior of large samples. See,
C. A. Stafford, A. J. Millis, and B. S. Shastry,
Phys. Rev. B {\bf 43}, 13660 (1991) and
R. M. Fye, M. J. Martins, D. J. Scalapino, J. Wagner, and W. Hanke,
Phys. Rev. B {\bf 44}, 6909 (1991).

\bibitem{puntilla}
Notice that recovering uncorrect Oguri's result within our more
general scheme is straightforward: it suffices the use of the second--order
selfenergy of the cluster instead the exact one that we use.

\bibitem{spinless1}
R. A. Molina, D. Weinmann, R. A. Jalabert, G-L. Ingold, and J-L. Pichard,
Phys. Rev. B {\bf 67}, 235306 (2003).

\bibitem{spinless2}
V. Meden and U. Schollw\"ock,
Phys. Rev. B {\bf 67}, 193303 (2003).

\bibitem{futuro}
We are presently conducting a deeper study of the interacting
spinless model aimed to answer the scaling behavior of the model as a
function of the number of {\it interacting} sites.
Conductances at different filling ratios will also be investigated.

\bibitem{pega}
Strictly speaking, the differential conductance obtained by our approach
and the many--body transmission probability given by Oguri in
Ref.(\onlinecite{oguri2}) are different physical magnitudes that can
differ. Nevertheless, we believe that in the absence of a source--drain
voltage dependence of the groundstate of the finite cluster, our energy
dependent $G$ is basically scanning the transmission of the states within the
allowed energy window. In particular, both magnitudes converge to 1 for
all states within the non--interacting band when $U$ vanishes.

\end{thebibliography}
\end{document}